\documentclass[10pt]{article} 
\usepackage{amsfonts,amssymb,slashed,makeidx,latexsym,setspace}
\usepackage{graphicx,graphics,floatflt,amssymb,epsf,rotate,subfigure} 
\usepackage{times,axodraw,cite,color}
\def\ltap{\raisebox{-.4ex}{\rlap{$\sim$}} \raisebox{.4ex}{$<$}}

\def\lprod{\lambda^{\prime *}_{223} \lambda^\prime_{323}}

\def\lamp{\lambda^\prime}
\def\twtwth{\lambda^\prime_{223}}
\def\thtwth{\lambda^\prime_{323}}

\begin{document} 
\begin{flushright} 
SINP/TNP/2009/25  
\end{flushright} 
 
\vskip 30pt 
 
\begin{center} 

  {\Large \bf Correlated enhancements in $D_s \to \ell\nu$, $(g-2)$ of muon,
    and lepton flavor violating $\tau$ decays with two $R$-parity violating
    couplings} \\

\vspace*{1cm} \renewcommand{\thefootnote}{\fnsymbol{footnote}} { {\sf Gautam
      Bhattacharyya ${}^1$}, {\sf Kalyan Brata Chatterjee ${}^{1}$}, and {\sf
      Soumitra Nandi ${}^{2}$}
} \\
  \vspace{10pt} {\small ${}^{1)}$ {\em Saha Institute of Nuclear Physics,
      1/AF Bidhan Nagar, Kolkata 700064, India} \\

    ${}^{2)}$ {\em Dipartimento di Fisica Teorica, Univ. di Torino and INFN,
       Sezione di Torino, I-10125 Torino, Italy}}\\
    
\normalsize 
\end{center} 

\begin{abstract}  

  With just two $R$-parity violating couplings, $\lambda'_{223}$ and
  $\lambda'_{323}$, we correlate several channels, namely, $D_s \to \ell \nu$
  ($\ell = \mu, \tau$), $(g-2)_\mu$, and some lepton flavor violating $\tau$
  decays. For $\lambda'_{223} = \lambda'_{323} \sim 0.3$ and for a common
  superpartner mass of 300 GeV, which explain the recently observed excesses
  in the above $D_s$ decay channels, we predict the following $R$-parity
  violating contributions: ${\rm Br} (\tau \to \mu \gamma) \sim 4.5 \cdot
  10^{-8}$, ${\rm Br} (\tau \to \mu\mu\mu) \sim 1.2 \cdot 10^{-8}$, ${\rm Br}
  (\tau \to \mu\eta/\eta') \sim 4 \cdot 10^{-10}$, and $(g_\mu-2)/2 \sim 4
  \cdot 10^{-11}$. We exhibit our results through observable versus observable
  correlation plots.

\vskip 5pt \noindent 
\texttt{PACS Nos:~~12.60.Jv, 13.35.Dx  } \\ 
\texttt{Key Words:~~$R$-parity violation, Lepton flavor violation}
\end{abstract} 

\setcounter{footnote}{0} 
\renewcommand{\thefootnote}{\arabic{footnote}}

\section{Motivation}
While all attention is now focused on the Large Hadron Collider (LHC) as a
possible {\em gold mine} of physics beyond the standard model (SM), one should
not lose sight of other territories rich with new physics, e.g., lepton flavor
violating (LFV) rare decays, which could provide complementary information.
Ever since neutrino flavor mixing was established, interests for observing
flavor violation in charged lepton decays have boomed. While in the neutrino
sector flavor violation could be rather large (maximal between $\nu_\mu$ and
$\nu_\tau$), in the charged lepton decays there is no sign of flavor violation
as yet.  The SM contributions to charged LFV decays are quite small, orders of
magnitude below the current experimental sensitivity, due to the smallness of
neutrino mass. Hence, any observation of LFV processes in the charged lepton
sector, which are being probed with ever increasing sensitivity, would
unambiguously point to non-standard interactions. Indeed, such indirect
observations taken in isolation may not imply much on the exact nature of
new physics.  But a study of possible correlations of its effects on different
independently measured charged LFV observables might provide a powerful
cross-check and lead to identification of new physics through LHC/LFV synergy.
In this paper, we consider $R$-parity violating (RPV) supersymmetry
\cite{rpar} and perform a correlation analysis of its numerical impact on
different LFV $\tau$ decays. We also study at tandem the RPV contribution to
$(g-2)_\mu$, an observable which continues to provide a $3\sigma$ room for new
physics despite significantly improved theoretical and experimental
accuracies.

$R$-parity is a discrete symmetry, which is defined as $R=(-1)^{3B+L+2S}$,
where $B$, $L$, and $S$ are the baryon number, lepton number and spin of a
particle, respectively. $R$ is $1$ for all SM particles and $-1$ for their
superpartners.  The usual assumption of $B$ and $L$ conservation in
supersymmetric models are not supported by any deep underlying principle. The
$L$-violating $\lamp$-type superpotential is written as ${\cal W}=
\lambda'_{ijk} L_i Q_j D^c_k$, where $L_i$ stands for SU(2) doublet lepton
superfields, $Q_i$ for SU(2) doublet quark superfields, $D^c_i$ for SU(2)
singlet down-type quark superfields, and $\{i,j,k\}$ are generation
indices. There are 27 such $\lamp$ couplings, on each of which and also on
many of their combinations exist strong constraints
\cite{reviews,Kao:2009fg}. We select {\em only two} of them, namely $\twtwth$
and $\thtwth$, and consider only them to be large and the rest to be either
vanishing or negligibly small.

{\em Why $\twtwth$ and $\thtwth$}? It has been observed that these two
couplings (each with a magnitude of $\sim 0.5$ and for superparticle masses
around 300 GeV) can justify the recently observed large $D_s \to \ell \nu$
$(\ell=\mu,\tau)$ branching ratios that the SM cannot explain
\cite{Kundu:2008ui}. On top of that, if we turn on even a small
$\lambda^\prime_{212} (\sim 0.001)$, then together with $\twtwth (\sim 0.5)$,
one can {\em also} explain the large phase in $B_s$--$\bar{B}_s$ mixing
\cite{Kundu:2008ui,Nandi:2006qe}. Turning our attention to the neutrino
sector, we recall that generation of neutrino masses and mixing by
$\lamp$-type couplings usually require a specific combination of indices,
namely, the $\lamp_{ijk}\lamp_{i'kj}$ product couplings \cite{gb}. But if we
take $\twtwth$ and $\thtwth$ (together with $\lamp_{113})$ as the only
non-vanishing and large RPV couplings, a phenomenologically acceptable pattern
of neutrino masses and mixing emerge at two-loop level \cite{Dey:2008ht}.  To
sum up, we consider four non-vanishing RPV couplings: two of them $\twtwth$
and $\thtwth$ large ($\sim 0.5$) and relevant for the present analysis, the
other two $\lambda^\prime_{212} \sim 0.001$ and $\lambda^\prime_{113}~\ltap~
0.1$ (from charged-current universality \cite{reviews}) not to be used in the
present analysis but implicitly present to justify our choices of $\twtwth$
and $\thtwth$ through the correlated phenomena mentioned above.  Now,
motivated by the observation of maximal mixing between $\nu_\mu$ and
$\nu_\tau$, we make a further assumption $\twtwth = \thtwth$. Keeping all
these in mind, we outline our agenda as follows: consider $\twtwth$ and
$\thtwth$ as the only two {\em relevant} RPV couplings (the other two,
viz.~$\lambda^\prime_{212}$ and $\lambda^\prime_{113}$, optional and small),
assume them to be real, set their magnitudes equal and just enough to explain
the $D_s \to \ell \nu$ anomaly, predict its effect on $(g-2)_\mu$, and
estimate its numerical impact on different LFV $\tau$ decays ($\tau \to
\mu\mu\mu$, $\tau \to \mu \gamma$, $\tau \to \mu \eta$, $\tau \to \mu
\eta^\prime$). For simplicity, we assume all squark and slepton masses to be
degenerate, and denote the common mass by $\tilde{m}$. We derive various
one-loop effective flavor violating vertices, which we have often referred to
as {\em form-factors}.  We display their exact as well as approximate
expressions. While for numerical plots we use the exact formulae, the
approximate expressions serve to provide an intuitive feel of the numerical
impact.

\section{$D_s \to \ell \nu$ and the $f_{D_s}$ anomaly}
The branching fraction of the leptonic decay $D_s \to \ell \nu$ ($\ell=\mu,
\tau$) is given by 
\begin{equation} 
\label{dlnu} 
{\rm Br} (D_s \to \ell \nu) = \frac{m_{D_s}}{8\pi} \tau_{D_s} f_{D_s}^2 
G_F^2 m_\ell^2 \left|V_{cs}\right|^2 \left(1 -
  \frac{m_\ell^2}{m_{D_s}^2}\right)^2 \, ,  
\end{equation}
where $\tau_{D_s}$ is the lifetime of $D_s$. The decay constant is defined as
$\langle 0\left|\bar{s} \gamma_\mu \gamma_5 c\right|D_s\rangle = i f_{D_s}
p_\mu$, where $p_\mu$ is the momentum of $D_s$. The branching ratio has a
helicity suppression factor characterized by $m_\ell^2$ on account of a
spin-zero particle decaying into two spin-half particles.  Monte-Carlo
simulations of QCD on lattice predict $f_{D_s} = 241 \pm 3$ MeV
\cite{Onogi:2009vc}. The experimental average is somewhat higher: $f_{D_s} =
277 \pm 9$ MeV \cite{pdg,:2008sq,:2007ws}. The enhancements are $(13 \pm 6)\%$
in the muon channel, $(18 \pm 8)\%$ in the tau channel, and $(15 \pm 5)\%$ on
average. On the other hand, the lattice estimate and the experimentally
obtained value for $f_D$ seem to be in perfect agreement around 206 MeV
\cite{Onogi:2009vc}. The latter suggests that the discrepancy in $f_{D_s}$ may
very well be influenced by new physics contributing in a {\em flavor specific
  way} to $D_s$ decay. Note that $D_s \to \ell \nu$ in the SM proceeds at tree
level and it is Cabibbo-allowed. Hence, loop suppressed new physics is an
unlikely candidate to account for the discrepancy.  Leptoquark or charged
Higgs interactions have been advocated in this context as they provide new
tree amplitudes for the above decay \cite{Dobrescu:2008er}.  Our candidate is
supersymmetric RPV interaction and our chosen couplings, $\twtwth$ and
$\thtwth$, contribute to $D_s \to \ell \nu_\ell ~(\ell=\mu,\tau)$ {\em via}
$\tilde{b}_R$-exchanged {\em tree} graphs \cite{Kundu:2008ui}. The net
contribution to the $D_s \to \mu \nu$ channel can be obtained by replacing
$G_F^2 \left|V_{cs}\right|^2$ in Eq.~(\ref{dlnu}) by
\begin{equation} 
\label{dsmunurpv}
  \left|G_F V_{cs}^\ast + \frac{\lambda^{\prime 2}_{223}}{4\sqrt{2} {\tilde
        m}^2} \right|^2 + 
     \left|\frac{\lambda^\prime_{223} \lambda^{\prime\ast}_{323}}
       {4\sqrt{2} {\tilde m}^2} \right|^2 \, . 
\end{equation} 
For $D_s \to \tau\nu$, we must do the replacements $\lambda^\prime_{223}
\leftrightarrow \lambda^\prime_{323}$ in Eq.~(\ref{dsmunurpv}).

\section{Anomalous magnetic moment of the muon}
The effective vertex of photon with any charged fermion is given by 
\begin{equation}
\bar{u}(p') \Gamma_\mu u(p) = \bar{u}(p') \left[\gamma_\mu F_1(q^2)
+ \frac{i\sigma_{\mu\nu} q^\nu}{2m_f} F_2(q^2) + \cdots\right] u(p) \, .
\end{equation} 
The muon magnetic moment for $f=\mu$ is given by $\vec{\mu} = g_\mu
\frac{e}{2m_\mu} \vec{s}$. At tree level, $F_1(0) =1$ and $F_2(0) =0$. Quantum
correction yields $a_\mu \equiv F_2(0) \ne 0$, while $F_1(0)$ remains unity at
all order due to charge conservation. Since $g_\mu \equiv 2\left(F_1(0) +
  F_2(0)\right)$, it follows that $a_\mu \equiv (g_\mu - 2)/2$.  As per 
current estimation \cite{gminus2}, the room for new physics is given by  
\begin{equation}
\label{g2num}
a_\mu^{\rm new} = a_\mu^{\rm exp} - a_\mu^{\rm SM} = (24.6 \pm 8.0) \cdot
10^{-10} \, .
\end{equation}
%%%%%%%%%%%%%%%%%%%%%%%%%%%%%%%%%%%%%%%%%%%%%
\begin{figure}[htbp]
\begin{center}
\resizebox{100mm}{!}{\includegraphics{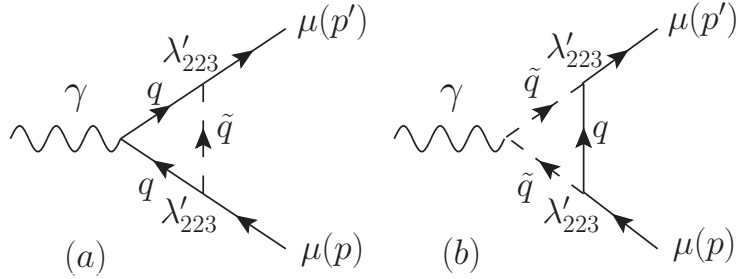}} 
\caption{\small{\sf $\lambda^\prime_{223}$-induced graphs contributing to
    $(g-2)_\mu$. Here, $(q, \tilde{q}) \equiv (c^c, \tilde{b}_R)$ and $(b,
    \tilde{c}_L^*)$.  }}
\label{gminustwo}
\end{center}
\end{figure}
%%%%%%%%%%%%%%%%%%%%%%%%%%%%%%%%%%%%%%%%%%%%%
The coupling $\lambda^\prime_{223}$ induces a contribution to $a_\mu$, which
proceeds through the diagrams in Fig.~\ref{gminustwo}. The quarks and squarks
inside the loop have been labeled by generic symbols $q$ and $\tilde q$
respectively, which can take two sets: $(q = c^c, \tilde q = \tilde{b}_R)$ and
$(q = b, \tilde q = \tilde{c}_L^*)$. The loop integrals would depend on $r_c
\equiv m_c^2/m_{\tilde{b}_R}^2$ and $r_b \equiv m_b^2/m_{\tilde{c}_L}^2$. As
mentioned earlier and assumed throughout our analysis, $m_{\tilde{b}_R} =
m_{\tilde{c}_L} = \tilde{m}$. We obtain
\begin{eqnarray}   
\label{amu_analytic}
a_\mu^{(\lambda^\prime)}&=& 3\frac{|\lambda'_{223}|^2 m_\mu^2}{16\pi^2{\tilde m}^2}
\left[\left\{Q_{c}\left(\xi_1(r_c)-\xi_2(r_c)\right)+
Q_{b}\left(\bar\xi_1(r_c)-\bar\xi_2(r_c)\right)\right\} \right. \nonumber \\
&-&\left. \left\{Q_b\left(\xi_1(r_b)-\xi_2(r_b)\right)+
Q_{c}\left(\bar\xi_1(r_b)-\bar\xi_2(r_b)\right)\right\}\right]
~\simeq ~ 3\frac{|\lambda'_{223}|^2 m_\mu^2}{16\pi^2{\tilde m}^2}
\left(\frac{1}{6}\right) \, .
\end{eqnarray}

The $\xi$-functions used throughout our analysis are given by 
\begin{equation}
\begin{array}{rclcl}
\xi_n(r)
   & = & \displaystyle \int_0^1 \frac{z^{n + 1} {\rm d} z}{1 + (r - 1) z}
   & = & \displaystyle \frac{-1}{(1-r)^{n+2} }
           \left[ \ln r + \sum_{k = 1}^{n + 1}
                            (-1)^{k} \pmatrix{n + 1 \cr k \cr}
                                                \frac{r^k -1 }{k}
           \right]
     \\[2ex]
\bar{\xi}_n(r) & = & \displaystyle {1\over{r}}
\xi_n\left({1\over r}\right)\ . &&
\end{array}
\label{xi}
\end{equation}

\section{$\tau^-\to\mu^-\mu^-\mu^+$ decay}
\begin{figure}[htbp]
\begin{center}
\resizebox{140mm}{!}{\includegraphics{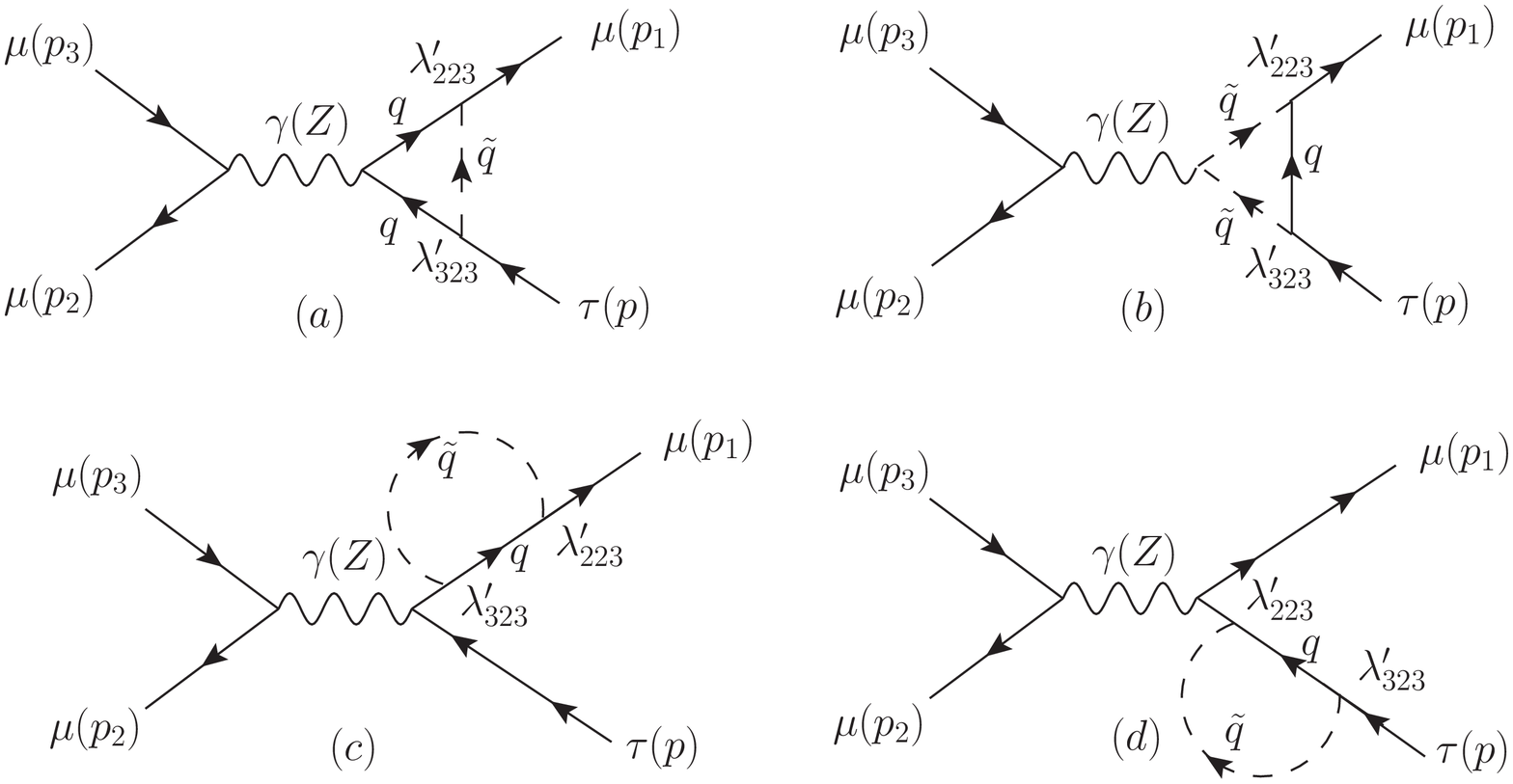}} 
\caption{\small{\sf $\lprod$-induced photon and $Z$-boson mediated penguins
    for $\tau^-\to\mu ^-\mu^-\mu^+$. Here, $(q, \tilde{q}) \equiv (c^c,
    \tilde{b}_R)$ and $(b, \tilde{c}_L^*)$.} }
\label{peng_tau3mu}
\end{center}
\end{figure}

\begin{figure}[htbp]
\begin{center}
\resizebox{140mm}{!}{\includegraphics{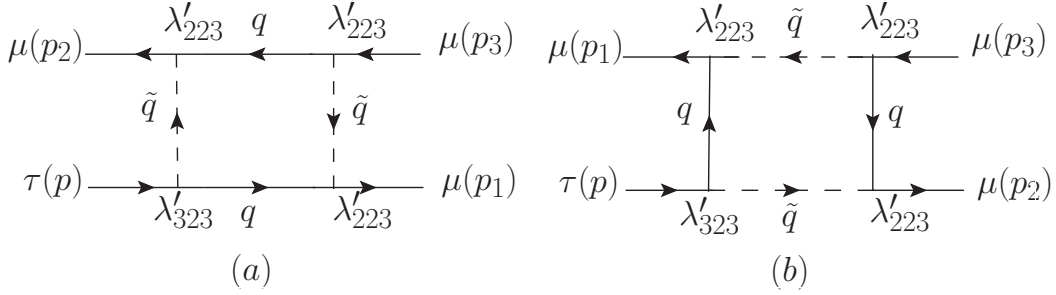}} 
\caption{\small{\sf Box graphs for $\tau^-\to\mu^-\mu^-\mu^+$.  Here, $(q,
    \tilde{q}) \equiv (c^c, \tilde{b}_R)$ and $(b, \tilde{c}_L^*)$. }}
\label{box_tau3mu}
\end{center}
\end{figure}

The decay $\tau \to \mu^-\mu^+\mu^-$ proceeds through photon and $Z$ penguins
(Fig.~\ref{peng_tau3mu}) and box graph (Fig.~\ref{box_tau3mu}). We consider
each of them below. Here flavor violation is induced by $\twtwth$ and
$\thtwth$ {\em via} loops with quarks and squarks in internal lines.

\subsection{Photon penguin}
%%%%%%%%%%%%%%%%%%%%%%%%%%%%%%%
%%%%%%%%%%%%%%%%%%%%%%%%%%%%%%%
The amplitude of the photon exchanged diagrams for $\tau^-\to\mu ^-\mu^-\mu^+$ 
decay can be written as
\begin{equation}
\label{mgamma}
M_\gamma = \bar u_\mu(p_1)\left[A_L q^2 \gamma_\mu P_L + 
i A_R m_\tau \sigma_{\mu\nu} q^\nu P_R\right]u_\tau(p)
\frac{e^2}{q^2}\bar u_\mu(p_2)\gamma^\mu v_\mu(p_3)-(p_1\leftrightarrow p_2)
\, , 
\end{equation}
where $q$ is the photon momentum. The form-factors $A_L$ and $A_R$ are induced
by the flavor-changing $\lprod$ couplings. Each penguin diagram will have a
quark ($q$) and a squark ($\tilde q$) inside the loop. There are two such
sets: $(q = c^c, \tilde q = \tilde{b}_R)$ and $(q = b, \tilde q =
\tilde{c}_L^*)$. We obtain
\begin{eqnarray}
A_L &=& \frac{3\lprod}{16\pi^2} 
\left(\frac{-2}{9 \tilde{m}^2} \right)
\left[\xi_2(r_c) + \frac{1}{2} \xi_2(r_b) + \frac{1}{4} \bar{\xi}_2(r_c)
+  \frac{1}{2} \bar{\xi}_2(r_b) \right]  \nonumber \\
&\simeq& \frac{3\lprod}{16\pi^2}
\left(\frac{1}{9 \tilde{m}^2} \right)
\left[5 + 4 \ln\left(\frac{m_c}{\tilde{m}} \right) 
+ 2 \ln\left(\frac{m_b}{\tilde{m}} \right)  \right] \, .
\end{eqnarray} 
The magnetic form-factor is given by 
\begin{eqnarray}
\label{ar}
 A_R&=& 3\frac{\lprod}{32\pi^2{\tilde m}^2}
\left[\left\{Q_{c}\left(\xi_1(r_c)-\xi_2(r_c)\right)+
Q_{b}\left(\bar\xi_1(r_c)-\bar\xi_2(r_c)\right)\right\} \right. \nonumber \\
&-&\left. \left\{Q_b\left(\xi_1(r_b)-\xi_2(r_b)\right)+
Q_{c}\left(\bar\xi_1(r_b)-\bar\xi_2(r_b)\right)\right\}\right]
~\simeq ~ 3\frac{\lprod}{32\pi^2{\tilde m}^2}
\left(\frac{1}{6}\right) \, .
\end{eqnarray}

\subsection{$Z$-boson penguin}
The $Z$-mediated penguin amplitude for the process $\tau^-\to\mu^-
\mu^-\mu^+$  is given by 
\begin{equation}
M_Z  =\bar u_\mu(p_1)\left[F_L \gamma_\mu P_L\right]u_\tau(p)
\frac{1}{M_Z^2} \bar u_\mu(p_2)\left[\gamma^\mu
  (a_L^\ell P_L+a_R^\ell P_R)v_\mu(p_3)\right] -
(p_1\leftrightarrow p_2) \, .
\end{equation}
The $Z$ boson couplings with the left- and right-chiral fermions are given by
\begin{equation}
\label{zcoup}
a_L^f= \frac{g}{\cos\theta_W}
\left(t_3^f - Q_f \sin^2\theta_W\right) \, , ~~~~
a_R^f= \frac{g}{\cos\theta_W} \left(- Q_f \sin^2\theta_W\right) \, . 
\end{equation}
The $\lprod$-induced contribution to the form-factor $F_L$ proceeds through two
sets of penguins: $(q = c^c, \tilde q = \tilde{b}_R)$ and $(q = b, \tilde q =
\tilde{c}_L^*)$, yielding  
\begin{eqnarray}
\label{fl}
F_L=\frac{g}{\cos\theta_W}
\left(\frac{3\lprod}{32\pi^2}\right)
\left[r_c \xi_0(r_c) - r_b \xi_0(r_b)\right] 
\simeq \frac{g}{\cos\theta_W}
\left(\frac{3\lambda'^*_{223}\lambda'_{323}}{32\pi^2}\right)
\left[\frac{m_b^2}{\tilde{m}^2} \left(1+2\ln\frac{m_b}{\tilde{m}}\right)
\right] \, .
\end{eqnarray}

%%%%%%%%%%%%%%%%%%%%%%%%%%%%%%%%%%%%%%%%%%%%%%%%%%%%%%%%%%%%%%%%%%%%%%%
\subsection{Box contribution}
The $\lambda^\prime_{223}$ and $\lambda^\prime_{323}$ couplings also induce a
box graph for $\tau^- \to \mu^- \mu^- \mu^+$ with internal quark and squark
lines. Again, two sets of box diagrams contribute $(q = c^c, \tilde q =
\tilde{b}_R)$ and $(q = b, \tilde q = \tilde{c}_L^*)$. The amplitude is given
by
%%%%%%%%%%%%%%%%%%%%%%%%%%%%%%%%%%%%%%%%%%%%%%%%%%%%%%%%%%%%%%%%%%%%%%
%%%%%%%%%%%%%%%%%%%%%%%%%%%%%%%%%%%%%%%%%%%%%%%%%%%%%%%%%%%%%%%%%%%%%%%
\begin{equation}
M_{\rm box}=e^2 B_L \left[\bar u_\mu(p_1)\gamma^\mu P_Lu_\tau(p)\right]
\left[\bar u_\mu(p_2)\gamma_\mu P_Lv_\mu(p_3)\right]-(p_1\leftrightarrow p_2)
\, . 
\end{equation}
For the sake of convenience, we normalize $B_L$ with a
prefactor $e^2$, though no gauge interaction is actually involved:  
\begin{equation}
\label{bl}
e^2 B_L = \frac{3\lambda^{\prime *2}_{223}\lambda'_{223}\lambda'_{323}}
{64\pi^2 \tilde{m}^2} \left[f(r_c)+ f(r_b)\right] \, , ~{\rm where}~
f(r)= \frac{1-r^2+2r\ln(r)}{(1-r)^3} \, . 
\end{equation}

\subsection{The branching ratio}
The total decay amplitude of this process is the sum of the penguin and box
contributions, given by $M_{\rm tot} = M_\gamma + M_Z + M_{\rm box}$. The
branching ratio of $\tau^-\to\mu^-\mu^-\mu^+$ is given in terms of the
different form-factors \cite{Arganda:2005ji}:
\begin{eqnarray}
\label{3lwidth}
{\rm Br} (\tau^-\to\mu^-\mu^-\mu^+) &=& 
\frac{e^4 m_\tau^5}{512\pi^3\Gamma_\tau} \left[A_L^2 - 4 A_L A_R
+ A_R^2 \left(\frac{16}{3} \ln \frac{m_\tau}{m_\mu}-\frac{22}{3}\right) 
 +\frac{1}{6} B_L^2 + \frac{2}{3} A_L B_L -\frac{4}{3} A_R B_L \right.
\nonumber \\
&+& \left. \frac{1}{3} \left(2F_{LL}^2 + F_{LR}^2 + 2B_LF_{LL}
+ 4 A_L F_{LL} + 2 A_L F_{LR} 
- 8 A_R F_{LL} - 4 A_R F_{LR}\right)\right] \, , 
\end{eqnarray}
where $\Gamma_\tau$ is the total decay width of $\tau$.  Our form-factors
($A_L, A_R, B_L, F_{LL}, F_{LR}$) are all real. The expressions of $F_{LL}$
and $F_{LR}$ are given by,
\begin{equation}
F_{LL} = \frac{F_La_L^\ell}{g^2 \sin^2\theta_W M_Z^2} \, ,
\hskip 30pt 
F_{LR} = \frac{F_L a_R^\ell}{g^2 \sin^2\theta_W M_Z^2} \, . 
\end{equation}

\section{Radiative decay $\mathbf {\tau\to\mu\gamma}$}
%%%%%%%%%%%%%%%%%%%%%%%%%%%%%%%%%%%%%%%%%%%%%%%%%%%%%%%%%%%%%%%%%%%%%%%
\begin{figure}[htbp]
\begin{center}
\resizebox{100mm}{!}{\includegraphics{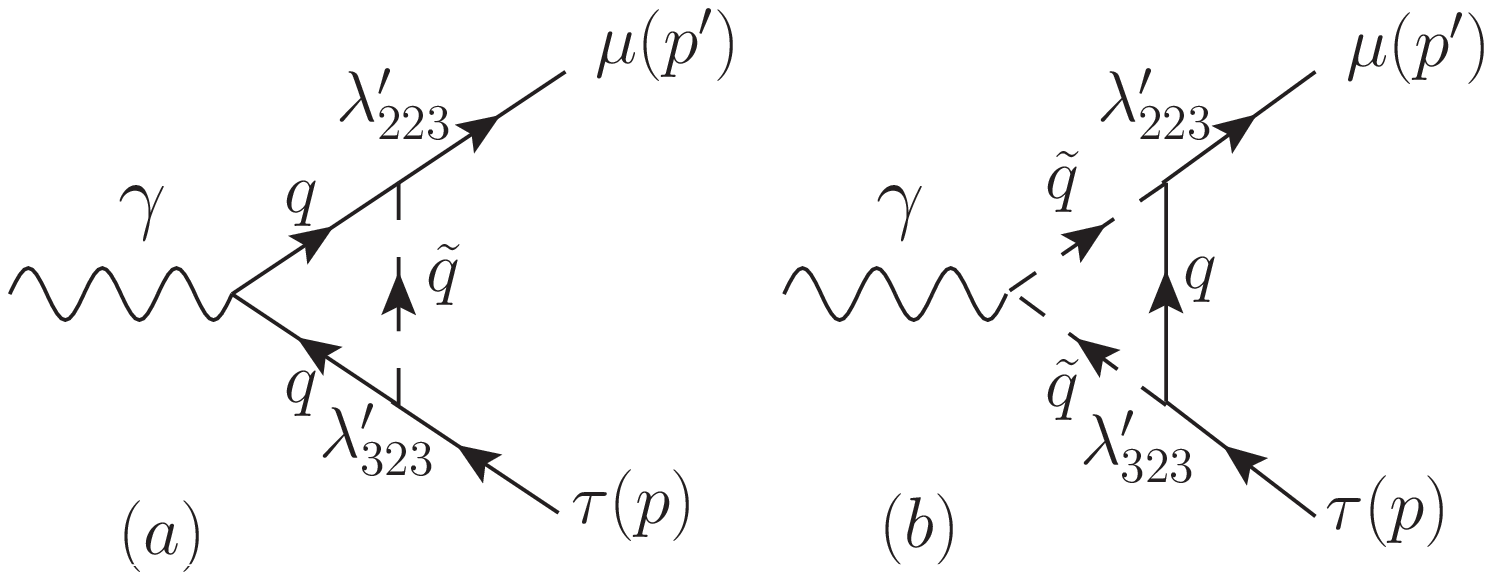}} 
\caption{\small{\sf $\lprod$-induced magnetic transition $\tau\to\mu\gamma$.
    Here, $(q, \tilde{q}) \equiv (c^c, \tilde{b}_R)$ and $(b,
    \tilde{c}_L^*)$.}}
\label{taumug}
\end{center}
\end{figure}
%%%%%%%%%%%%%%%%%%%%%%%%%%%%%%%%%%%%%%%%%%%%%%%%%%%%%%%%%%%%%%%%%%%%%%
We have shown in Fig.~\ref{taumug} how $\twtwth$ together with $\thtwth$
drive the magnetic transition $\tau \to \mu \gamma$. The amplitude for this
transition is given by 
\begin{equation} 
M(\tau \to \mu \gamma) = A_R m_\tau \bar{u}_\mu(p') 
(i\sigma_{\mu\nu} q^\nu P_R) u_\tau(p) \epsilon^{\mu\ast} \, ,  
\end{equation} 
where $\epsilon^\mu$ is the photon polarization. The expression for $A_R$ can
be found in Eq.~(\ref{ar}). In the amplitude we have neglected a similar term
proportional to $m_\mu$. The branching ratio for this radiative decay mode is
given by (neglecting any $m_\mu$-dependent term)
\begin{equation}
{\rm Br} (\tau^-\to\mu^-\gamma)=
\frac{e^2}{16\pi \Gamma_\tau}m_\tau^5 A_R^2 \, .
\end{equation}

\section{Semileptonic lepton flavor violating $\tau$ decay
: $\tau\to\mu\eta(\eta')$}
\begin{figure}[htbp]
\begin{center}
\resizebox{140mm}{!}{\includegraphics{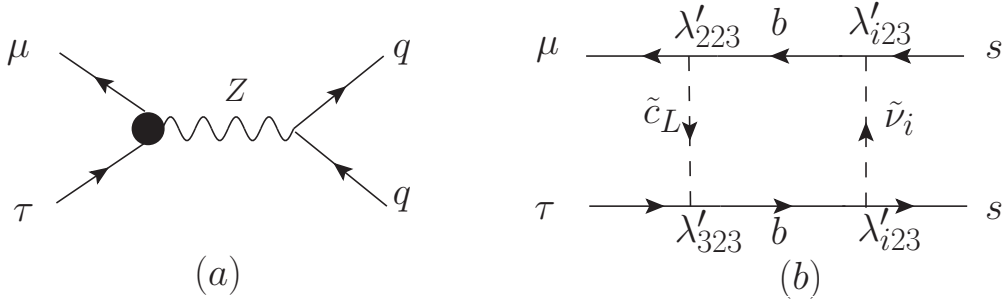}} 
\caption{\small{\sf The penguin and box amplitudes for $\tau\to\mu P$, where
    $P=\eta,\eta'$. The quark level diagrams are shown, which proceed through
    (a) $Z$ penguin, and (b) box graph. The blob in (a) is a symbolic
    representation of lepton flavor violating vertex as in
    Fig.~\ref{peng_tau3mu}. In (a), we take $q=u,d,s$, all of which contribute
    to the formation of $\eta$ and $\eta^\prime$. In (b), the specific indices
    of $\lamp$ couplings ensure only $s\bar{s}$ final state is produced. The
    index $i$ of $\lambda^\prime_{i23}$ takes two values: $i=2,3$. There is
    another box diagram, not drawn above, where both the internal scalars are
    $\tilde{b}_R$ and the internal fermions are $c$ and $\nu_i$. }}
\label{taumup}
\end{center}
\end{figure}

The semileptonic decay $\tau\to\mu P$ with $P=\eta(\eta')$ decay is mediated
by a $Z$-penguin and a box graph, as shown in Figs.~(\ref{taumup}a) and
(\ref{taumup}b), respectively.  Photon penguin cannot contribute as it cannot
provide the axial current for the quarks to condense to a meson.

The $Z$-boson mediated penguin amplitude for $\tau \to \mu q \bar{q}$ is given
by
\begin{equation}
\label{zpenguin-mup}
M_Z (\tau \to \mu q \bar{q}) =
\bar{u}_\mu \left[\gamma_\mu F_LP_L \right] u_\tau
\frac{1}{M_Z^2}\bar{u}_q\left[\gamma^\mu (a^q_L P_L+a^q_R P_R)\right]v_q \, , 
\end{equation}
where $a^q_L$ and $a^q_R$ are given in Eq.~(\ref{zcoup}). The relevant $q$
for the formation of $\eta$ and $\eta^\prime$ are $u,d$ and $s$. The form
factor $F_L$ is already given in Eq.~(\ref{fl}).

The couplings $\twtwth$ and $\thtwth$ also induce $\tau \to \mu q \bar{q}$
through box graphs. Because of the specific $\lamp$-indices, $q$ can only be
$s$. The box graph contains two fermion lines and two scalar lines. There are
two types of box diagrams: (i) the fermions are the same ($b$ quark), but the
scalars are different 
($\tilde{c}_L$ and $\tilde{\nu}_{\mu L}/\tilde{\nu}_{\tau L}$); 
(ii) the scalars are same ($\tilde{b}_R$), but
the fermions are different ($c$ and $\nu_\mu/\nu_\tau$).  The sum of box
amplitudes is given by,
\begin{equation}
\label{box-mup}
  M_{\rm box} (\tau \to \mu s \bar{s}) = 
D_L \left[\bar{u}_\mu\gamma^\mu P_L u_\tau\right] 
\left[\bar{u}_s\gamma_\mu P_L v_s\right]
\, . 
\end{equation}
The form-factor $D_L$ is given by 
\begin{equation}
\label{dl}
D_L = {3 \over {64 \pi^2 \tilde{m}^2}} \sum_{i=2,3} 
\lambda^{\prime}_{323}\lambda^{\prime *}_{223} 
\left|\lambda^{\prime}_{i23}\right|^2
\left[f(r_b) + f^\prime(r_c)\right] \simeq 
{3 \over {32 \pi^2 \tilde{m}^2}} \sum_{i=2,3} 
\lambda^{\prime}_{323}\lambda^{\prime *}_{223} 
\left|\lambda^{\prime}_{i23}\right|^2 \, , 
\end{equation}
where $f(r)$ has already been expressed in Eq.~(\ref{bl}), while $f^\prime(r)$
is given by
\begin{equation}
\label{fprime}
f^\prime(r)= \frac{1-r+r\ln(r)}{(1-r)^2} \, . 
\end{equation}
Using Eqs.~(\ref{zpenguin-mup}-\ref{fprime}) we obtain the branching ratio,
\begin{equation}
\label{brtaumup}
{\rm Br} (\tau\to\mu P) \simeq 
\frac{(m_\tau^2 - m_P^2)^2}{16\pi m_\tau \Gamma_\tau}\Big(
\Big|{D_L\over 2}+{g\over {2 \cos{\theta_W}}} t^{s}_3 {F_L\over
  {M^2_Z}}\Big|^2 {f^{s}_{P}}^2 + 2 \Big|{g\over {2 \cos{\theta_W}}} t^{u/d}_3 
{F_L\over{M^2_Z}}\Big|^2 {f^{u}_{P}}^2\Big) \, .  
\end{equation}
The decay constants involving $\eta$ and $\eta^\prime$ are given by 
\begin{eqnarray}
f^{u}_{\eta} = f^{d}_{\eta} &=& {1\over \sqrt{6}} f_8 \cos\theta_8 -
{1\over \sqrt{3}} f_0\sin\theta_0 \, , ~~~
f^{u}_{\eta^{\prime}} = f^{d}_{\eta^{\prime}} = {1\over \sqrt{6}} f_8 \sin
\theta_8 +{1\over\sqrt{3}} f_0 \cos\theta_0 \, , \nonumber \\
f^{s}_{\eta} &=& -{2\over \sqrt{6}} f_8 \cos\theta_8 -{1\over \sqrt{3}} f_0
\sin\theta_0 \, , ~~~
f^{s}_{\eta^{\prime}} =-{2\over \sqrt{6}} f_8 \sin\theta_8 +{1\over
  \sqrt{3}} f_0 \cos\theta_0  \, .
\end{eqnarray}
The numerical values of the involved parameters are given by
\cite{Feldmann:1997vc,pdg};
\begin{equation}
f_8 = 168 ~{\rm MeV}, ~~ f_0 = 157 ~{\rm  MeV}, ~~ \theta_8 = - {22.2}^{\circ}, 
~~\theta_0 = - {9.1}^{\circ}, ~~ m_\eta~(m_{\eta^\prime}) = 547.8~(957.7)~{\rm
  MeV} \, . 
\end{equation}

\section{Results} 
In Table 1 we have displayed the present experimental status of
different branching ratios of our concern. 
\begin{table}[htbp]
\begin{center}
\begin{tabular}{|c||c|}
\hline
 Decay modes & Branching fractions    \\
\hline \hline
$D_s^+\to \mu^+\nu_\mu$  & $(6.3 \pm 0.5)\times 10^{-3}$ \\
\hline
$D_s^+\to \tau^+\nu_\tau$ &  $(6.6 \pm 0.6)\times 10^{-2}$    \\
\hline
$\tau^-\to\mu^-\mu^-\mu^+$ & $< 3.2\times 10^{-8}$ \\
\hline
$\tau^-\to\mu^-\gamma$ & $< 4.5\times 10^{-8}$ \\
\hline
$\tau^-\to\mu^-\eta$ & $< 6.5\times 10^{-8}$ \\
\hline
$\tau^-\to\mu^-\eta^\prime$ & $< 1.3\times 10^{-7}$ \\
\hline
\end{tabular}
\caption{\small{\sf Present status of the observed branching ratios of
    $D_s^+\to \ell^+\nu$ ($\ell = \mu, \tau$) and the experimental upper
    limits on different LFV $\tau$ decays at 90\% C.L. We have quoted numbers
    cited in Particle Data Group \cite{pdg}, although slightly stronger
    constraints in some channels exist \cite{Banerjee:2007rj}. The expected
    reach at the Super$B$ factory with 75 ${\rm ab}^{-1}$ data for $\tau^- \to
    \mu^- \mu^- \mu^+$ and $\tau^- \to \mu^- \eta$ channels are $2 \cdot
    10^{-10}$ and $4 \cdot 10^{-10}$, respectively -- see the Super$B$
    conceptual design report \cite{Bona:2007qt}.}}
\label{tabfb}
\end{center}
\end{table}

\noindent {\em Existing limits on $\lamp$}: We reiterate that all our
processes are driven by $\twtwth$ and $\thtwth$.  The existing limits on them
depend on $m_{\tilde{b}_R}$. As mentioned before, throughout our analysis we
have assumed a common sparticle mass of 300 GeV.  The best upper limit on
$\twtwth$ comes from $R_{D^0}\equiv \frac{{\rm Br} (D^0 \to K^- \mu^+
  \nu_\mu)}{{\rm Br} (D^0 \to K^- e^+ \nu_e)}$, and the limit is 0.3 at 90\%
C.L.~\cite{Kao:2009fg,Bhattacharyya:1995pq}\footnote{It is interesting to
  observe that the 2$\sigma$ upper limit $\left|\lambda^\prime_{22k}\right| <
  0.16$ obtained in \cite{Bhattacharyya:1995pq} with $R_{D^0} = 0.84 \pm 0.12$
  is not much different from the latest update
  $\left|\lambda^\prime_{22k}\right| < 0.1$ at 2$\sigma$ for $\tilde{m} = 100$
  GeV using $R_{D^0} = 0.92 \pm 0.04$ \cite{Kao:2009fg}. In spite of a
  significant reduction of the error on $R_{D^0}$ (over a period of 14 years),
  the 2$\sigma$ upper limit on $\left|\lambda^\prime_{22k}\right|$ remained
  more or less the same because the central value gradually moved towards
  unity.}.  On the other hand, the best upper limit on $\thtwth$ arises from
$R_{D_s}(\tau\mu) \equiv \frac{{\rm Br} (D_s^+ \to \tau^+ \nu_\tau)}{{\rm Br}
  (D_s^+ \to \mu^+ \nu_\mu)}$, the limit being 0.9 at 90\%
C.L.~\cite{Kao:2009fg}.

\noindent {\em Our parameters}: Recall that maximal mixing between $\nu_\mu$
and $\nu_\tau$ motivated us to assume $\twtwth = \thtwth = \lamp$.  For
showing numerical correlations through different plots, we scan $\lamp$ in the
range $[0 - 0.7]$, but keep $\tilde{m}$ fixed at 300 GeV. Also, the band width
in each plot is a consequence of varying $f_{D_s}$ by 2$\sigma$ around its
central value, i.e. in the range $[235-247]$ MeV.  As we go from left to right
in each band the value of $f_{D_s}$ increases.  The minimum values of $\lamp$
consistent with 2$\sigma$ lower limits of ${\rm Br} (D_s\to\mu\nu)$ and ${\rm
  Br} (D_s\to\tau\nu)$ are 0.3 and 0.4, respectively, which correspond to (the
2$\sigma$ upper limit of) $f_{D_s} = 247$ MeV.

\begin{figure}[htbp]
\begin{center}
\resizebox{65mm}{!}{\includegraphics{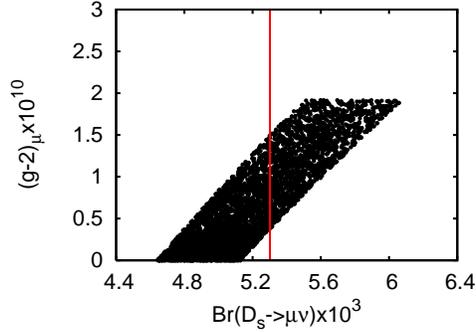}} 
\caption{\small{\sf{Correlation between the ${\rm Br} (D_s\to\mu\nu)$ and the
      muon anomalous magnetic moment. The $\twtwth$-induced contribution to
      the latter is well below the current experimental sensitivity -- see
      Eq.~(\ref{g2num}). The vertical line indicates 2$\sigma$ lower limit of
      the branching ratio -- see Table 1. } }}
\label{plot_g2}
\end{center}
\end{figure}

\noindent {\em Contribution to $(g-2)_\mu$}: Following
Eq.~(\ref{amu_analytic}), we obtain
\begin{equation} 
\label{amu_numeric}
a_\mu^{\lamp} \simeq 1.9 \cdot 10^{-10} \left(\frac{\lamp}{0.7}\right)^2 
\left(\frac{300}{\tilde{m} ~{\rm (in~GeV)}} \right)^2 \, . 
\end{equation}
In Fig.~\ref{plot_g2}, we have plotted the correlation between contribution to
$(g-2)_\mu$ along one axis and the branching ratio of $D_s\to\mu\nu$ along the
other. We note here that the $R$-parity conserving contribution to $(g-2)_\mu$
can be sizable too for large $\tan\beta$. In fact, an approximate expression
for $\tan\beta \gg 1$ can be found in \cite{Carena:1996qa} as
\begin{equation} 
\label{amu_mssm}
a_\mu^{\rm MSSM} \simeq 1.7 \cdot 10^{-10} \tan\beta 
\left(\frac{300}{\tilde{m} ~{\rm (in~GeV)}} \right)^2 \, . 
\end{equation}

\begin{figure}[htbp]
\begin{center}
\begin{tabular}{cc}
\resizebox{65mm}{!}{\includegraphics{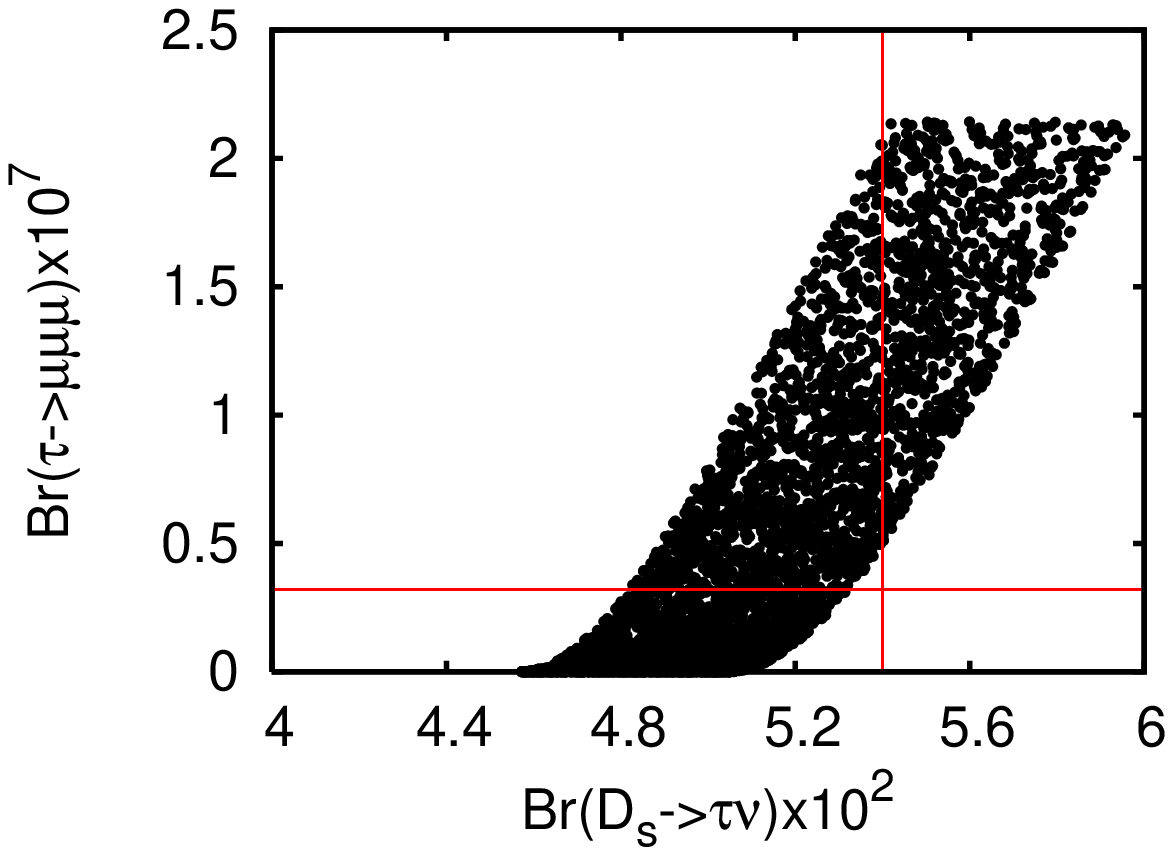}} &
\resizebox{65mm}{!}{\includegraphics{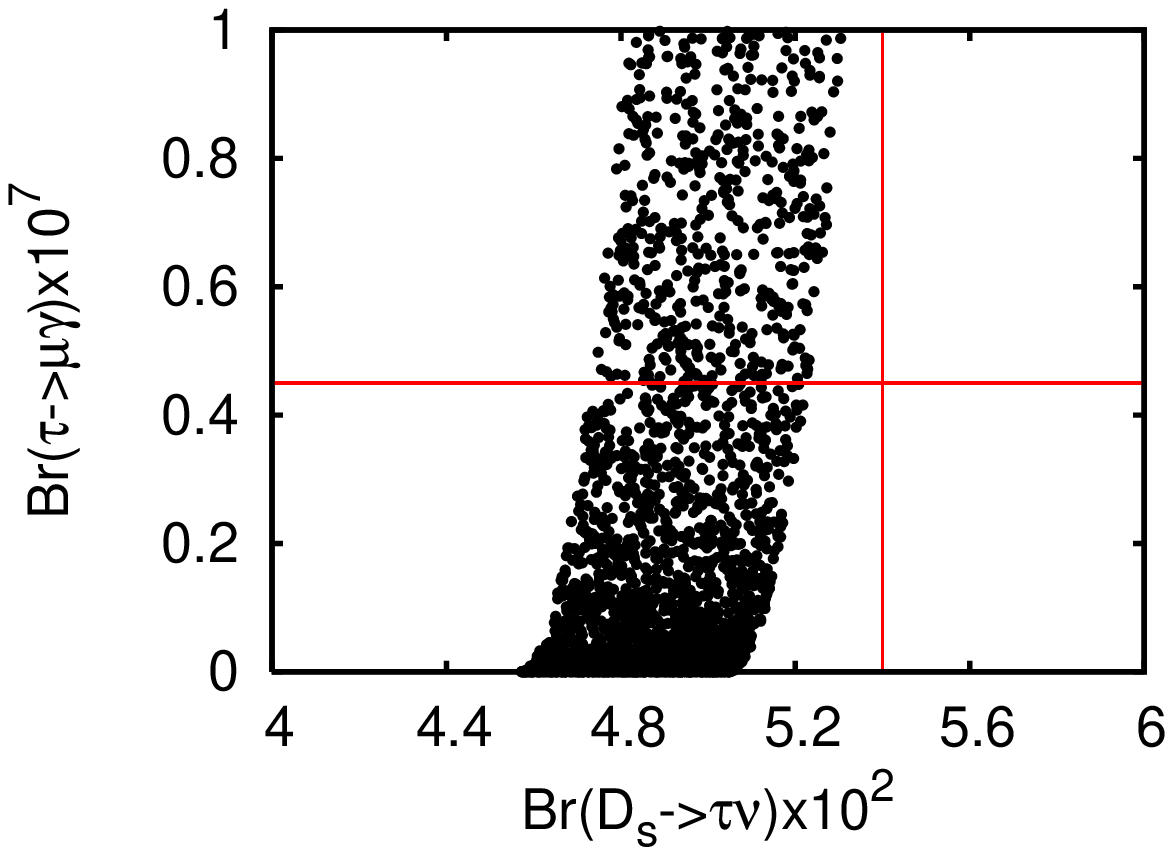}}\\
\end{tabular}
\caption{\small{\sf Correlations between (a) ${\rm Br} (\tau\to\mu\mu\mu)$ and
    ${\rm Br} (D_s\to\tau\nu)$ in the left panel, and (b) ${\rm
      Br} (\tau\to\mu\gamma)$ and ${\rm Br} (D_s\to\tau\nu)$ in the right
    panel. The vertical line in each plot corresponds to 2$\sigma$ lower limit
    of the $D_s$ branching ratio, while the horizontal lines are the 90\%
    C.L. upper limits of the LFV $\tau$ decay branching ratios -- see Table
    1.}}
\label{plot_mug3mu}
\end{center}
\end{figure}

\noindent {\em LFV $\tau$ decays}: The approximate (yet, to a very good
accuracy) expressions of the LFV form-factors with their explicit dependence
on $\lamp$ and $\tilde{m}$ (the other parameters are all known) are as
follows:
\begin{eqnarray} 
\label{formfactors}
A_L \simeq - 2.9 \cdot 10^{-7} ~{\rm GeV}^{-2} 
\left(\frac{\lamp}{0.7}\right)^2 
\left(\frac{300}{\tilde{m} ~{\rm (in~GeV)}} \right)^2 \, , \,\,\,\,
A_R \simeq 8.6 \cdot 10^{-9} ~{\rm GeV}^{-2} 
\left(\frac{\lamp}{0.7}\right)^2
\left(\frac{300}{\tilde{m} ~{\rm (in~GeV)}} \right)^2  \, , \nonumber \\
B_L \simeq 2.8 \cdot 10^{-7} ~{\rm GeV}^{-2} 
\left(\frac{\lamp}{0.7}\right)^4 
\left(\frac{300}{\tilde{m} ~{\rm (in~GeV)}} \right)^2 \, , \,\,\,\,
F_{LL} \simeq 1.3 \cdot 10^{-9} ~{\rm GeV}^{-2} 
\left(\frac{\lamp}{0.7}\right)^2 
\left(\frac{300}{\tilde{m} ~{\rm (in~GeV)}} \right)^2 \, ,  \\
F_{LR} \simeq -1.1 \cdot 10^{-9} ~{\rm GeV}^{-2} 
\left(\frac{\lamp}{0.7}\right)^2 
\left(\frac{300}{\tilde{m} ~{\rm (in~GeV)}} \right)^2  \, , \,\,\,\, 
D_L \simeq 5.1 \cdot 10^{-8} ~{\rm GeV}^{-2} 
\left(\frac{\lamp}{0.7}\right)^4 
\left(\frac{300}{\tilde{m} ~{\rm (in~GeV)}} \right)^2  \, . \nonumber
\end{eqnarray}
Since each box has four $\lamp$-vertices, $B_L$ and $D_L$ have both quartic
sensitivity to $\lamp$, while the penguin form-factors have quadratic
dependence on $\lamp$. Using the expressions in Eq.~(\ref{formfactors}), we
calculate the branching ratios to a very good approximation as 
\begin{eqnarray} 
\label{eq:t3mu}
{\rm Br} (\tau \to \mu\mu\mu) &\simeq& 3.9 \cdot 10^{-7} 
\left[1.0 - 0.6 \left(\frac{\lamp}{0.7}\right)^2 +
0.1 \left(\frac{\lamp}{0.7}\right)^4\right] 
\left(\frac{\lamp}{0.7}\right)^4 
\left(\frac{300}{\tilde{m} ~{\rm (in~GeV)}}\right)^4 \, ; \\
\label{eq:tmug}
{\rm Br} (\tau \to \mu\gamma) &\simeq& 1.0 \cdot 10^{-6} 
\left(\frac{\lamp}{0.7}\right)^4 
\left(\frac{300}{\tilde{m} ~{\rm (in~GeV)}}\right)^4 \, ; \\
\label{eq:tmueta}
{\rm Br} (\tau \to \mu\eta) &\simeq& 3.4 \cdot 10^{-7} 
\left(\frac{\lamp}{0.7}\right)^8 
\left(\frac{300}{\tilde{m} ~{\rm (in~GeV)}}\right)^4 \, ; \\
\label{eq:tmuetap}
{\rm Br} (\tau \to \mu\eta^\prime) &\simeq& 3.3 \cdot 10^{-7} 
\left(\frac{\lamp}{0.7}\right)^8 
\left(\frac{300}{\tilde{m} ~{\rm (in~GeV)}}\right)^4 \, .
\end{eqnarray} 
In Eq.~(\ref{eq:t3mu}), the first term within the square bracket is the pure
penguin contribution, the second term represents interference between penguin
and box graphs, while the last term is the pure box contribution. As explained
before, the $\lamp$ dependence is different for different terms. Also, by
comparing Eq.~(\ref{eq:tmug}) with Eq.~(\ref{eq:t3mu}), we observe that for
the same choices of $\lamp$ and $\tilde{m}$ the prediction of ${\rm Br} (\tau
\to \mu \gamma)$ is one order of magnitude enhanced compared to ${\rm Br}
(\tau \to \mu \mu \mu)$. This happens primarily because the latter is a 3-body
decay which involves more suppression factors which cannot compensate the fact
that $\left|A_R\right| \simeq \left|A_L\right|/33$.  Figs.~\ref{plot_mug3mu}a
and \ref{plot_mug3mu}b capture the numerical correlations. We observe that the
region allowed at 2$\sigma$ by $D_s \to \tau\nu$ overshoots the 90\%
C.L. upper limit of the branching ratio of $\tau \to \mu\mu\mu$.  Obviously,
the same thing happens for $\tau \to \mu \gamma$.  However, we should keep in
mind that the branching ratio of $D_s \to \ell \nu$ ($\ell = \tau$ in the
present context) has not only an experimental uncertainty, but also inherits a
theoretical uncertainty from $f_{D_s}$. Even by {\em mild} stretching of one
or both of these uncertainties beyond 2$\sigma$, it is possible to accommodate
both $\tau \to \mu\mu\mu$ and $\tau \to \mu\gamma$.

\noindent {\em New Limit}: The upper limit on $\tau \to \mu \gamma$ branching
ratio (see Table 1) restricts $\lamp < 0.3$. Our limit is stronger than before
and, without any need of the assumption $\twtwth = \thtwth$, should be
interpreted as
\begin{equation}
\label{newlimit} 
\left|\lambda_{223}^{\prime\ast} \thtwth \right| < 0.09 ~~{\rm at~ 90\% ~C.L.}
\end{equation}
Since the $D_s$ branching ratios require $\lamp > (0.3-0.4)$ at 2$\sigma$ (or
90\% C.L.), to avoid conflict with Eq.~(\ref{newlimit}) we need to stretch the
present limits, as already mentioned. If we fix $\lamp = 0.3$, the prediction
for the branching ratio of $\tau \to \mu\mu\mu$ is $\sim 1.2 \cdot 10^{-8}$,
which is roughly a factor of 3 below the current sensitivity, but still very
much within the reach of the super$B$ factory with 75 ${\rm ab}^{-1}$
projected luminosity. If, however, $\tau \to \mu \gamma$ remains elusive even
at super$B$, then as per our prediction, $\tau\to\mu\mu\mu$ is not going to be
observed either.

\begin{figure}[htbp]
\begin{center}
\begin{tabular}{cc}
\resizebox{65mm}{!}{\includegraphics{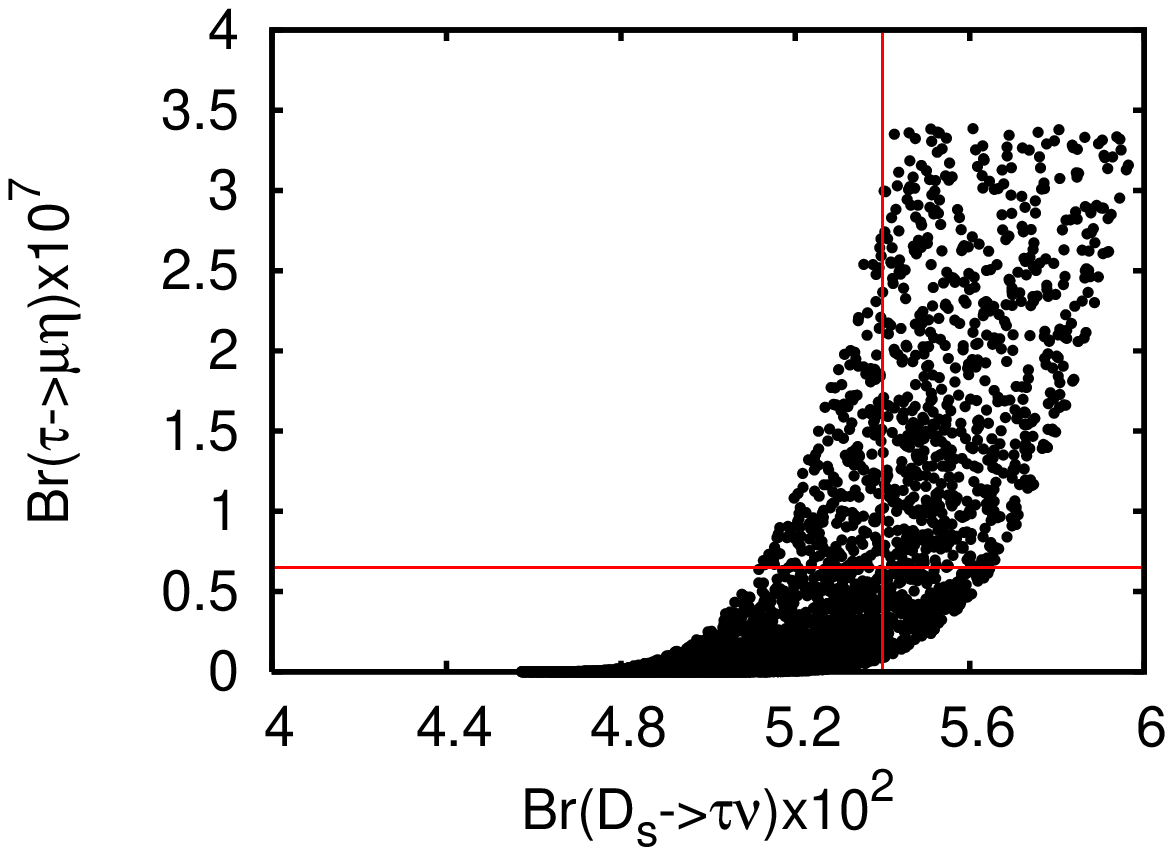}} &
\resizebox{65mm}{!}{\includegraphics{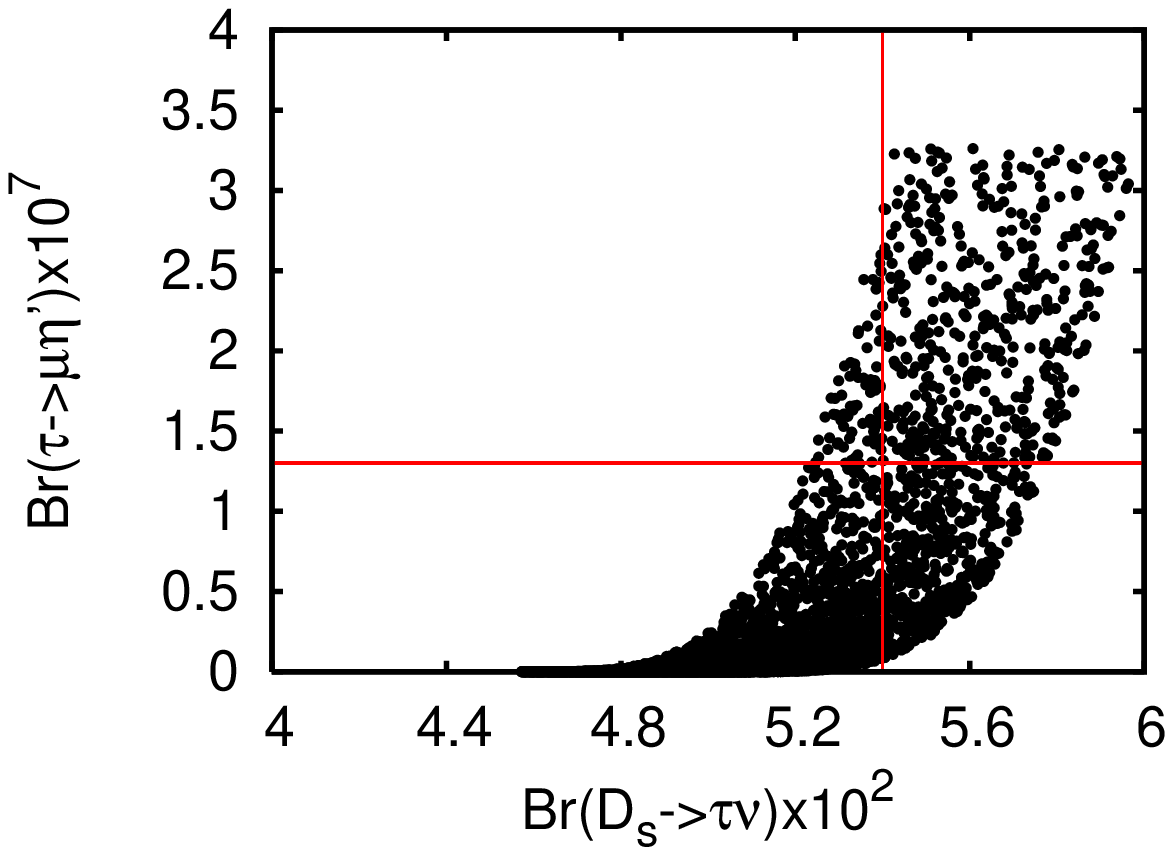}}\\
\end{tabular}
\caption{\small{\sf Exactly like in Fig.~\ref{plot_mug3mu}, except that the
    branching ratios of $\tau \to \mu \eta$ and $\tau \to \mu \eta^\prime$ are
    plotted in (a) the left panel and (b) the right panel, respectively. The
    horizontal lines are the 90\% C.L. upper limits on the branching ratios.}}
\label{plot_etaetap}
\end{center}
\end{figure}

The branching ratio of $\tau \to \mu P$, expressed in Eq.~(\ref{brtaumup}),
contains contributions from box (the $D_L$ part) and penguin (the $F_L$
part). The box contribution is significantly more dominant than the
penguin. We display the approximate numerical values of the branching ratios
for $P = \eta$ and $\eta^\prime$ in Eqs.~(\ref{eq:tmueta}) and
(\ref{eq:tmuetap}), respectively.  The dependences on $\lamp$ and $\tilde{m}$
are similar as both processes involve similar box graphs. As expected, these
modes are not as constraining as $\tau \to \mu \gamma$. If we put $\lamp =
0.3$, the branching ratios for the $\eta$ and $\eta^\prime$ modes are
predicted to be around $4 \cdot 10^{-10}$, i.e. two orders of magnitude below
the present sensitivity, but within the accuracy expected to be reached at the
super$B$ factory with 75 ${\rm ab}^{-1}$ luminosity. Again, a positive signal
at Super$B$ necessarily requires an observation of $\tau\to\mu\gamma$ at the
current sensitivity. The numerical correlations of $\tau \to \mu \eta$ and
$\tau \to \mu \eta^\prime$ decay modes with the $D_s \to \tau \nu$ branching
fraction have been plotted in Fig.~\ref{plot_etaetap}.

\section{Comparison with previous works and conclusions} 
We divide this section in three parts: (i) we briefly mention about the
existing studies on $R$-parity conserving supersymmetric contribution to LFV
$\tau$ decays, (ii) remark on the previous works on $R$-parity violating
contributions to lepton flavor violation, and finally (iii) highlight the new
things that we have done in this work.

\noindent (i)~ LFV decays have been analyzed in supersymmetric scenarios with
conserved $R$-parity but with different sets of supersymmetry breaking
parameters. In a class of scenarios where minimal supersymmetry is augmented
by three right-handed neutrino superfields for generating neutrino masses {\em
  via} see-saw mechanism, it has been shown
\cite{Arganda:2005ji,Arganda:2008jj} that large neutrino Yukawa couplings
induce large flavor violation in the slepton sector which is ultimately
transmitted to the LFV observables. The general conclusion is that light
supersymmetry ($m_0, M_{1/2} < 250$ GeV) is disfavored. Large LFV branching
ratios (with large $\tan\beta \sim 50$) can be obtained when light neutrino
masses are hierarchical. In general, $\tau \to \mu\gamma$ is the most
sensitive LFV channel, but to explore the Higgs sector $\tau \to \mu \eta$ and
$\tau \to \mu \eta^\prime$ channels are more effective.  It has been shown
that in a general {\em unconstrained} minimal supersymmetric framework
\cite{Brignole:2004ah}, for low $\tan\beta \sim 3$, the branching ratio in the
$\tau\to \mu\mu\mu$ channel is ${\cal{O}} (10^{-9})$ and in the $\tau \to \mu
\eta (\eta^\prime)$ channel less than $10^{-10}$. On the other hand, for large
$\tan\beta \sim 50$ and for small pseudo-scalar mass ($m_A$), the Higgs
mediated contributions are extremely dominant. In the latter case, indeed with
strong fine-tuning of parameters, ${\rm Br} (\tau \to \mu\mu\mu)$ is enhanced
to ${\cal{O}} (10^{-7})$ and ${\rm Br} (\tau \to \mu\eta)$ to even larger
values.  In supersymmetric models embedded in minimal SO(10) group
\cite{Fukuyama:2005bh}, the LFV branching ratios are, however, several orders
of magnitude below the present experimental sensitivities.

\noindent (ii)~ RPV induced LFV processes have been studied in the past in
different contexts \cite{lfv-rpv}.  Except $\ell_i \to \ell_j \gamma$, all
other LFV processes considered there proceed at {\em tree level} with
appropriately chosen RPV couplings. The choices of such couplings are, in
general, different in different processes.  Their primary intentions were to
put upper limits on different single and product couplings by confronting LFV
observables with experimental results.

\noindent (iii)~ {\em What are the new things that we have done in this
  paper}? We made an economical choice of RPV couplings ($\twtwth$ and
$\thtwth$ only), motivated {\em primarily} by their ability to explain the
large $D_s \to \ell \nu$ ($\ell = \mu,\tau$) branching ratios.  We set these
two couplings equal, a choice inspired by maximal $\nu_\mu$-$\nu_\tau$
mixing. We have kept the sparticle mass fixed at 300 GeV. Explanation of $D_s
\to \mu \nu (\tau\nu)$ branching ratios require $\lamp > 0.3 (0.4)$ at 90\%
C.L. On the other hand, $\tau\to\mu\gamma$ with an upper limit of $4.5 \cdot
10^{-8}$ on its branching ratio at 90\% C.L. offers the most sensitive LFV
probe of the RPV dynamics, and sets {\em an improved upper limit} $\lamp <
0.3$ at 90\% C.L.  Enhanced theoretical and experimental accuracies in the
$D_s \to \ell \nu$ channels might eventually release the tension between the
apparently conflicting requirements. Putting $\lamp = 0.3$, we obtain ${\rm
  Br} (\tau \to \mu\mu\mu) \sim 1.2 \cdot 10^{-8}$, and ${\rm Br} (\tau \to
\mu\mu\mu) \div {\rm Br} (\tau \to \mu \eta/\eta^\prime) \simeq 30$.  The
correlation plots capture the underlying dynamics. To sum up, instead of
considering just one experimental observation at a time, be it an anomaly or
an excess {\em vis-\`{a}-vis} the SM expectation, providing a {\em raison
  d'\^{e}tre} for one set of new interactions, we have studied the possibility
of correlated enhancements in a variety of LFV channels using just two RPV
couplings. We demonstrated our results through `observable versus observable'
plots.

\noindent {\bf Acknowledgments:}~ GB thanks the CERN Theory Division for
hospitality and acknowledges a partial support through the project
No.~2007/37/9/BRNS of BRNS (DAE), India.  SN's work is supported by a European
Community's Marie-Curie Research Training Network under contract
MRTN-CT-2006-035505 `Tools and Precision Calculations for Physics Discoveries
at Colliders'.

\end{document}